\documentclass[12pt]{article}

\input{epsf}

\newcommand{\bea}{\begin{eqnarray}}
\newcommand{\eea}{\end{eqnarray}}
\newcommand{\be}{\begin{equation}}
\newcommand{\ee}{\end{equation}}
\newcommand{\pa}{\partial}
\newcommand{\nn}{\nonumber \\}

\newcommand{\tr}{\mbox{tr}}

\newcommand{\hphi}{\hat{\phi}}

\makeatletter

\@addtoreset{equation}{section}
\makeatother

\def\href#1#2{#2}

\begin{document}



\begin{center}
\ \vspace*{5mm}\\
{\bf \large Matrix Model For Polyakov Loops,%
\vspace*{1mm}\\
String Field Theory In The Temporal Gauge,%
\vspace*{1mm}\\
Winding String Condensation In Anti-de Sitter Space%
\vspace*{2mm}\\
And Field Theory Of D-branes}%
\vspace*{10mm}\\
{{\sc Furuuchi \ Kazuyuki}}\\
\vspace*{4mm}
{\sl National Center for Theoretical Sciences\\
National Tsing-Hua University, Hsinchu 30013, Taiwan, R.O.C.}\vspace*{0.5mm}\\
{\tt furuuchi@phys.cts.nthu.edu.tw}
\vspace*{3mm}\\
\begin{abstract}
\noindent Closed string field theory is constructed
by stochastically quantizing
a matrix model for Polyakov loops
that describes phases of a large $N$ 
gauge theory at finite temperature.
Coherent states in this string field theory
describes winding string condensation
which has been expected to cause 
a topology change from thermal AdS geometry
to AdS-Schwarzschild black hole geometry.
D-branes in this closed string field theory is also discussed.
Slightly extended version of a talk given at CosPA 2007, Nov.13-15, Taipei, Taiwan.
\vspace*{2mm}\\
\noindent {\it Keywords}: {AdS-CFT correspondence; stochastic quantization; 
loop equations; string field theory.}
\end{abstract}
\end{center}


\section{Introduction}

AdS-CFT correspondence \cite{Maldacena:1997re}
provides a way to understand
closed string theory which contains {quantum gravity} --
one of the most important but also the most challenging issues in
theoretical physics --
using dual non-gravitational field theory.
However, the correspondence still remains as a conjecture.
Therefore, it is important to investigate the dictionary 
between closed string theory and the dual field theory.
In this talk, I will explain my trial for
translating
gauge field theory language
into
closed string field theory language,
in a simplified setting.
The first part of this talk is based on 
the work Ref.~\cite{Furuuchi:2006st}.
In this section, I briefly
review the background materials.
For more detail, please consult my lecture note
Ref.~\cite{Furuuchi:2006pk} and references therein.

\subsection{Matrix model for Polyakov loops}

The setting I choose to carry out my program
is the AdS$_5$-CFT$_4$ correspondence --
namely, the duality between closed string on AdS$_5 \times S_5$
and ${\cal N}=4$ super Yang-Mills theory 
with $SU(N)$ gauge group -- at finite temperature.
I study the case where the Yang-Mills theory
lives on a spatial manifold $S^3$.
One of the expected phenomena for 
Yang-Mills theory at finite temperature
is the confinement-deconfinement phase transition.
Although the space on which the Yang-Mills theory lives on
is compact ($S^3$),
I will mostly consider 
the limit $N \rightarrow \infty$ where
the large $N$ phase transition can take place.

A class of order parameters for the confinement-deconfinement
transition which
I will focus on 
are expectation values of Polyakov loops
winding around the Euclidean time direction
for $n$ times:
\bea
{\cal P}_n
\equiv 
\frac{1}{N} \tr P \, e^{i\int_0^{n\beta} d\tau A_0} 
\label{Pn} ,
\eea
where $\tr$ is the trace over $SU(N)$ gauge indices
$P$ denotes the path ordering, and $\beta$
is the inverse temperature.
To determine the phase which is realized by calculating
the expectation values of these operators,
one can integrate out, in principle,
all the fields
except the constant mode of 
the temporal component of the gauge field \cite{Aharony:2003sx}:
\bea
\langle {\cal P}_n \rangle
=
\int [dA_0]\,
{\cal P}_n
e^{-S_{eff}(A_0)} .
\eea
In the following, I will use $A_0$ 
to denote the constant part.
Actually, 
$A_0$ is identified with $A_0 + 2\pi/\beta$ by a gauge transformation,
and it is suitable to use
the variable $U \equiv e^{i\beta A_0}$:
\bea
 \label{MMP}
\langle {\cal P}_n \rangle =
\int [dU]\, \frac{1}{N} \tr U^{n} \, e^{{-S'}_{eff}(\tr U^n)},
\eea
where ${S'}_{eff}$ follows from the change of variables.
Eq.~(\ref{MMP}) is what I call
``matrix model for Polyakov loops".
In practice, it is not easy to perform the path
integration of other fields explicitly.
However, there are general features
which the effective action $S'_{eff}(\tr U^n)$ should have.
In particular,
the fields in the ${\cal N}=4$ super Yang-Mills theory
are all in the adjoint representation of $SU(N)$,
and when compactified on the Euclidean time circle
the action has $Z_N$ symmetry.
It follows that the effective
action ${S'}_{eff}(\tr U^n)$ is also invariant under
the $Z_N$ action:
\bea
 \label{Zn}
U \rightarrow U\,  e^{i\frac{2\pi}{N} m} ,
\eea
with $m$ integer.\footnote{%
Actually, because of the $Z_N$ symmetry
the expectation value of a single Polyakov loop is zero.
When we use the large $N$ saddle point approximation,
a contribution from each saddle point is not zero,
and I will mostly discuss such contribution in this note.}
This is the property which 
characterizes the important physics of our model,
as I will explain in the next section.

\subsection{Gravitational phase transition in asymptotically AdS space}

Let us start from the classical gravity regime,
i.e. $N \rightarrow \infty$ with large 't Hooft
coupling. 
In this case, from the classical ``approximation"\footnote{Actually
we haven't constructed quantum (completion) of the gravitational
theory, so here we are assuming the existence of such theory.}
 of the
Euclidean path integral gravity,
one observes a phase transition at some temperature,
the low temperature phase is thermal AdS geometry
and high temperature phase is Euclidean
AdS-Schwarzschild black hole geometry
(Hawking-Page phase transition) \cite{Hawking:1982dh}.
\begin{figure}
\begin{center}
 \leavevmode
 \epsfxsize=70mm
 \epsfbox{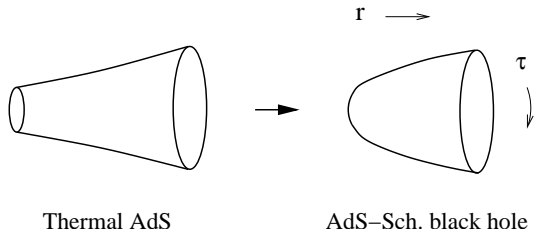}\\
\end{center}
\caption{Hawking-Page phase transition.
\label{topoc}}
\end{figure}
In Fig.~\ref{topoc}, the Euclidean time and the radial directions
are depicted.
This phase transition was identified
as the holographic dual
of the confinement-deconfinement phase transition
\cite{Witten:1998qj,Witten:1998zw}.
One of the evidences for this identification
comes from the dictionary of the AdS-CFT correspondence
for calculating the expectation value
of Wilson/Polyakov loops from the closed string side
\cite{Rey:1998ik,Maldacena:1998im}:
\bea
 \label{graP}
\langle {\cal P} \rangle
\sim
e^{-{T_{st}\cal A}}  ,
\eea
where 
${\cal A}$ is an area of string worldsheet in the
gravitational background which ends on the Polyakov loop
on the boundary, and $T_{st}$ is the string tension.
From the Eq.~(\ref{graP}), 
one observes that the difference of the 
topology of the target space
is reflected
in the expectation value of the Polyakov loops:
A string worldsheet with disk topology 
(the leading order in the $1/N$ expansion)
can wrap the AdS-Schwarzschild black hole geometry
but cannot wrap the thermal AdS geometry,
see Fig.~\ref{topoc}.
In the case of finite 't Hooft coupling,
Eq.~(\ref{graP}) should be replaced
with path integral for closed string action
in the gravitational back ground
with string $\alpha'$ corrections.
Precisely speaking,
in the AdS$_5$-CFT$_4$ correspondence
what is given by the right hand side
of Eq.~(\ref{graP})
is not the ordinary Wilson/Polyakov loop
expectation value, but the
generalization of it including the
adjoint scalars in 
${\cal N}=4$ super Yang-Mills theory \cite{Maldacena:1998im}.
Since we could consistently integrate out
these scalars in the gauge theory side to obtain
(\ref{MMP}),
it would be reasonable to
assume that we can also consistently
integrate out the degrees of freedom 
in the closed string side
to obtain a consistent closed string
theory in two dimension.
Then, the dictionary Eq.~(\ref{graP}), 
or with appropriate string corrections 
in the right hand side, will give
expectation values of the 
Polyakov loops of our matrix model.

\subsection{Winding string condensation?}

In the case of string theory in a flat space, 
when a compactified circle shrinks to the 
size near the string scale
the winding mode of the string becomes tachyonic 
(in the case of superstring, this happens 
when fermions are anti-periodic
in the circle direction).
It is tempting to interpret the 
Hawking-Page phase transition as
a result of winding string condensation:
\begin{figure}
\begin{center}
 \leavevmode
 \epsfxsize=90mm
 \epsfbox{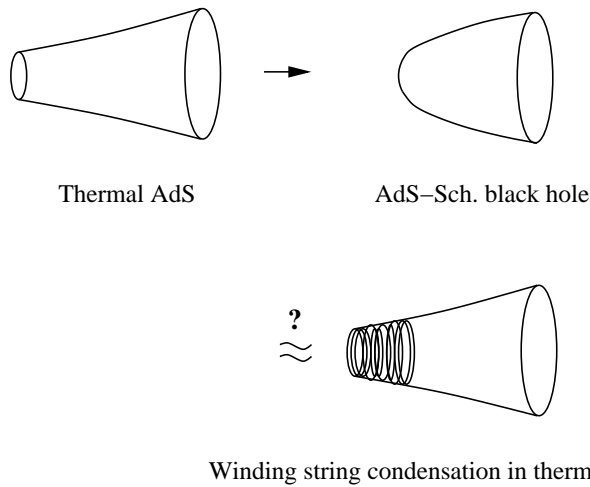}\\
\end{center}
\caption{Is Hawking-Page phase transition winding string condensation?
\label{wcond}}
\end{figure}
In the case of open string, tachyon condensation
describes the annihilation of D-brane
\cite{Sen:1998sm,Sen:1999nx,Schnabl:2005gv}.
One can expect that similarly
when winding strings condensate,\footnote{%
Having a tachyonic mode is not necessary for
condensation.}
it causes a disappearance of space-time.
The topology change from the thermal AdS geometry
to the AdS-Schwarzschild black hole geometry
may be understood as a result of this kind of
winding string condensation (Fig.~\ref{wcond})
\cite{KalyanaRama:1998cb,Barbon:2001di,Barbon:2002nw,Barbon:2004dd,%
McGreevy:2005ci,Horowitz:2006mr}.
In order to describe such condensation,
field theory would be the most suitable formalism.
However, since 
closed string field theories 
are quite complicated,
and currently known formulation 
are not manifestly background independent
(see e.g. Refs.~\cite{Sen:1993mh,Sen:1993kb,Zwiebach:1996ph}), 
it has been quite hard to 
describe condensation of string field 
and the change of the background.
However,
in the AdS-CFT correspondence the background independence
of the closed string side is partially realized 
since if one fixes the boundary field theory,
only the asymptotic geometry is fixed:
States in the boundary theory correspond to
normalizable modes in the bulk.
Therefore, we can expect to learn something about
background independence
if we can construct closed string field theory
in the AdS-CFT correspondence.
Fortunately, in the present simplified setting
there is a formalism of closed string field theory available,
which I will explain in the next section.

\section{String field theory in the temporal gauge}

String field theory in the temporal gauge
is a way to construct a closed string field theory from
given one-matrix model \cite{Ishibashi:1993pc}.
The name follows from the temporal gauge fixing on
the worldsheet of non-critical string \cite{Kawai:1993cj,Fukuma:1993tp}.
A good review of the subject is Ref.~\cite{Ikehara:1994xs}.

In Ref.~\cite{Jevicki:1993rr}, the temporal gauge quantization
was identified with the stochastic quantization 
plus
the change of variables from matrix to loops.
Let us first review the procedure of the stochastic quantization
(see e.g. Ref.~\cite{Damgaard:1987rr} for a review).
To extract the essential point,
I take a scalar field theory in 0 dimension 
in the path integral formalism, i.e. 
ordinary integral, as an example.
In this case, the expectation value of
observable $F(\phi)$ is given by
\bea
 \label{PI0}
&&\langle
F(\phi)
\rangle
=
{\cal N'}
\int d\phi \, e^{-S(\phi)} F(\phi), \nn 
&&{{\cal N'}^{-1} = \int d\phi \, e^{-S(\phi)}} .
\eea
The stochastic quantization is a way to
obtain the weight $e^{-S(\phi)}$ for the path integral
as a result of stochastic processes.
For this purpose, we prepare 
probability distribution $P(t;\phi)$:
\bea
\int d\phi \, P(t;\phi) = 1 ,
\eea
which satisfies the Fokker-Planck equation:
\bea
 \label{FPeq}
\frac{\pa}{\pa t} P(t;\phi)
=
-\frac{\pa}{\pa \phi}
\left(
\frac{\pa}{\pa \phi} +\frac{\pa S}{\pa \phi} 
\right) P(t;\phi)
\equiv
-H_{FP} \, P(t;\phi) .
\eea
The solution of Eq.~(\ref{FPeq}) is given by
\bea
 \label{formal}
P(t;\phi) = e^{-t H_{FP}} 
P(0;\phi) .
\eea
After taking the 
``stochastic time" 
$t \rightarrow \infty$,
we obtain the path integral weight as a stationary configuration:
\bea
 \label{limit}
\lim_{t\rightarrow \infty} P(t;\phi) = e^{-S(\phi)} .
\eea
The temporal gauge quantization can be 
phrased as a
rewriting of the above formula 
by the operator formalism.
Let us introduce 
``creation" and ``annihilation" operators
$\hat{\phi}^{\dagger}$, $\hat{\phi}$:
\bea
[\hat{\phi},\hat{\phi}^{\dagger}]=1 .
\eea
We have the following map:
\bea
 \label{map}
\phi \leftrightarrow \hat{\phi}^{\dagger}, &&
\frac{\pa}{\pa \phi}  \leftrightarrow \hat{\phi},\nn 
f(\phi) = \langle \phi| f(\hphi^{\dagger})|0\rangle,&&
\langle \phi | \hat{\phi}^{\dagger}
= 
\langle \phi | \phi , \quad
\phi | 0 \rangle = 0 .
\eea
Here, $\dagger$ has nothing to do with the Hermitian conjugation.
The relation between $\hat{\phi}^{\dagger}$ and $\hat{\phi}$ 
is rather that of coordinate and momentum.
This point will be different
in our model, as I will explain shortly.
 
Using the map (\ref{map}), we can rewrite
Eq.~(\ref{PI0}) as
\bea
&&\int d\phi \, e^{-S(\phi)} F(\phi)  \nn 
&&=
\lim_{t\rightarrow \infty}
\int d\phi \, 
\left(
e^{-t H_{FP}(\phi,\pa/\pa\phi)}
P(0;\phi)
\right) 
F(\phi) \nn 
&&=
\lim_{t\rightarrow \infty}
\int d\phi \,
\langle \phi | P(0;\hphi^\dagger) e^{-t \hat{H}_{FP}(\hphi^\dagger,\hphi)}
F(\hat{\phi}^{\dagger})
|0\rangle  .
\eea

By applying the temporal gauge quantization to
our matrix model of Polyakov loops (\ref{MMP}),
we obtain a closed string field theory.
The application of the stochastic quantization
plus change of variables to loops
as possible explanation of 
the AdS-CFT correspondence was
first discussed in 
Refs.~\cite{Periwal:2000dq,Lifschytz:1999fw,Lifschytz:2000bj}.
An important feature of our model is the $Z_N$ symmetry
Eq~(\ref{Zn}),
whose physical relevance is explained in the following.

I choose a map from path integral variables
to operators as follows:\footnote{%
Similar formulas have appeared in 2D Yang-Mills theory \cite{Minahan:1993np,Douglas:1993wy}, basically
because it also reduces to a unitary matrix model.}
\bea
 \label{Umap}
\tr U^n \leftrightarrow{\sqrt{n}} \, a_n^{\dagger} ,\quad
\tr U^{-n} \leftrightarrow {\sqrt{n}} \, \bar{a}_n^{\dagger} ,
\eea
where I have introduced
creation and annihilation operators for a string with winding number 
$n$ and $-n$ ($n > 0$), which satisfy
the following commutation relations:
\bea
[a_m, a_n^{\dagger}]= \delta_{mn}, \quad
[\bar{a}_m, \bar{a}_n^{\dagger}]= \delta_{mn} .
\eea
In the previous example of the temporal quantization,
the operators introduced were
more like position and momentum operators.
However, here
since $\tr U^n$'s are complex variables,
we can instead
use the usual creation and annihilation operators.
There is a freedom
in the choice of the overall normalization
in the map (\ref{Umap}),
which does not affect the final results.\footnote{However,
in the $N \rightarrow \infty$ limit
whether to include $1/N$ factor in the right hand sides
of (\ref{Umap}) makes difference
for the possible range of
eigenvalues of the coherent states bellow 
from infinite to finite, so one should be 
little careful in how to take the large $N$ limit.
The normalization should be carefully chosen in the
double scaling limit 
studied in Refs.~\cite{Alvarez-Gaume:2005fv,AlvarezGaume:2006jg}.
Here, I just chose the normalization so that
the fermionization formula discussed in the next
section becomes simple.
In this note I will not get into this point further
and leave the above subtle point to the future study.}

By rewriting the matrix model for Polyakov loop (\ref{MMP})
using the map (\ref{Umap}), we obtain
a closed string field theory
(see Ref.~\cite{Furuuchi:2006st} for the detail):
\bea
 \label{SFTT}
\langle {\cal P}_n \rangle &=&
\int [dU]\, \frac{1}{N} \tr U^{n} \, e^{{-S'}_{eff}(\tr U^n)} \nn
&=&
\frac{1}{N^2}
\lim_{t\rightarrow \infty}
\langle f |
e^{-t \hat{H}_{FP}(a_m,a_m^{\dagger},\bar{a}_m,\bar{a}_m^{\dagger})} 
a_n^{\dagger} 
|0\rangle ,
\eea
where the final state $\langle f |$ will be explained bellow.

The closed string field theory 
we have thus obtained is a peculiar type, 
in that it only contains winding modes.
It is asymmetric under T-duality transformation
since it does not contain momentum zero modes.
This is similar to the case studied in 
Ref.~\cite{Minahan:1993np,Douglas:1993wy}.
The appearance of the stochastic time
in the stochastic quantization is very
suggestive for the AdS-CFT correspondence.
In the AdS-CFT correspondence,
the dimension of the space-time in which 
closed strings live in 
is higher than that of the dual field theory,
and because of this reason
the AdS-CFT correspondence is regarded as 
a concrete realization of the idea called holography.
What is particularly important is 
the radial direction of the AdS space which is
related to the energy scale of the field theory.
The stochastic time has a scaling dimension,
and therefore it should be related to 
the radial direction.\footnote{%
In Ref.~\cite{Ishibashi:1993pc} 
the stochastic time was identified
with the worldsheet time
in a certain gauge \cite{Fukuma:1993tp}. 
We may expect a relation between
the target space time and the worldsheet time
after an appropriate gauge fixing.}
Whether this is a good way
to understand the relation between
the radial direction in the closed string side
and the energy scale in the dual field theory side
remains to be seen,
but I hope my work provides a support for this strategy.

The final state $\langle f |$ is specified by
the stationarity condition
\bea
 \label{stnl}
\langle f | \hat{H}_{FP} = 0  .
\eea
The condition Eq.~(\ref{stnl}) 
may not uniquely determine the final state.
Therefore, I will use the input from the 
matrix model.
Then, the final state 
is determined by the phase of the gauge theory,
which depends on the temperature.
To describe the final state, 
it is convenient to introduce
coherent states:
\bea
\langle w_n | a_n^{\dagger} =
\langle w_n | w_n .
\eea
Then, in the confined phase,
the final state is given by
\bea
\langle 0 | .
\eea
In the deconfined phase, it is given by
\bea
 \label{condense}
\langle w_n^c |  ,
\eea
where $w_n^c$ are the expectation value of 
the Polyakov loops calculated from the matrix model.
Eq.~(\ref{condense}) describes the
winding string condensation, as we wished at the
end of the last section.

In our matrix model, 
the action had the $Z_N$ symmetry (\ref{Zn}).
The Fokker-Planck Hamiltonian 
is also invariant under the $Z_N$ transformation.
This is translated into
the closed string field theory 
side as winding number conservation 
during the stochastic time evolution,
in the $N \rightarrow \infty$ limit.
However, in the deconfined phase,
the winding number conservation is 
violated by the condensation of the winding string,
since the coherent state is a superposition
of states with different winding numbers.
This is consistent with the {topology}
of the AdS-Schwarzschild black hole geometry
where the winding number is not a conserved quantity.

When $N$ is finite,
the winding number is conserved modulo $N$.
This is reminiscent to 
the stringy exclusion principle \cite{Maldacena:1998bw}
(see also Ref.~\cite{Lifschytz:1999fw}).
However, one should note that
when $N$ is finite, the notion of space-time
is not classical \cite{Maldacena:2001kr},
and hence the concept of 
winding number is not clear in the closed string side.
For finite $N$, one should also 
take into account 
the Madelstam relations among the loops.

In our formulation, the information of the 
background is encoded in the final state.
The Fokker-Planck Hamiltonian does not contain
the information of the condensation.
This should be compared with the
usual formulation of string field theory
\cite{Sen:1993mh,Sen:1993kb}.
See also Refs.~\cite{Ishibashi:1995in} for a different 
but related formulation.
If we consider higher dimensional case, for
example the
full AdS$_5$-CFT$_4$ correspondence,
the field theory calculations contain
divergences, and 
in particular the Fokker-Planck Hamiltonian
needs regularization.\footnote{%
For related studies of loop equations,
which is essentially
the Fokker-Planck Hamiltonian applied to the Wilson loops,
in the context of the AdS-CFT correspondence,
see for example Refs.~\cite{Hirano:1999ay,Hata:2005bk}.}
The information of the
background of the closed string,
which corresponds to the phase of the gauge theory,
can enter in the Hamiltonian if one uses a 
regularization which depends on the phase.

\section{Field theory of D-branes}

D-branes 
played crucial roles in uncovering the
non-perturbative aspects of string theory.
D-branes are often referred to as {solitons} 
in string theory.
In order to describe D-branes
as solitons, the most suitable framework
would be field theory of strings.
However, because
closed string field theory is technically quite complicated,
actual description of D-branes as solitons 
is quite limited at this moment.
Therefore, it will be useful to gain insights
by studying
D-branes in our simple closed string field theory.
In the following, I would like to construct
creation and annihilation operators
of D-branes.

The eigenvalues of gauge field compactified
on a circle
are identified with positions of D-branes
in the T-dual coordinate. 
Since our matrix model 
has multi-trace interaction
terms, the
eigenvalues interact among themselves,
as opposed to the familiar single trace matrix models
of non-critical strings.
However, for the purpose of identifying 
creation and annihilation operators of D-branes,
it would be enough to study
free D-branes, i.e.
the case where the potential term among eigenvalues
are absent.
This can be formally realized as zero temperature limit
of the matrix model obtained 
from integrating out modes other than the temporal 
component of the gauge field in
the free ${\cal N} =4$ super Yang-Mills theory.
In this case, the action reduces to 
the logarithm of the Vandermonde determinant.\footnote{This
may also look like interaction among eigenvalues,
but we normally regard it as just giving the Fermi
statistics.}
If we
pick out one eigenvalue from the rest,
it may be regarded as
position of a single D-brane.
This consideration leads to the conclusion that
adding one eigenvalue to the matrix model amounts to
inserting following factor
\bea
 \label{VdM}
\int_0^{\frac{2\pi}{\beta}} dx\,
\left|
\det 
\left\{
\sin \frac{\beta}{2} (x-A_0) 
\right\}
\right|^2
\eea
into the $SU(N-1)$ matrix model path integral, 
with $x$ being the eigenvalue for the single D-brane.
Eq.~(\ref{VdM}) can be rewritten as
\bea
\int_0^{\frac{2\pi}{\beta}}  dx\, \exp 
\sum_{n > 0}
\left(
\frac{1}{n} \tr U^n e^{-in\beta x}
+\frac{1}{n} \tr U^{-n} e^{in\beta x}
\right).
\eea
After the temporal gauge quantization,
this is expressed by putting 
the following operator to the initial state,
similar to what was found in Refs.~\cite{Hanada:2004im,Ishibashi:2007gh}
(see also \cite{Fukuma:1996hj,Fukuma:1996bq,Fukuma:1999tj})
in the case of Hermitian matrix models:
\bea
\int_0^{\frac{2\pi}{\beta}} dx \, \, \psi(x)\bar{\psi}(x) ,
\eea
where
\bea
 \label{fermi}
\psi(x) \equiv : e^{i \phi(x)} : \, , \quad 
\bar{\psi}(x) \equiv : e^{i \bar{\phi}(x)} : \, ,
\eea
and $:\, :$ denotes the appropriate
operator ordering
(see e.g. Ref.~\cite{Green:1987sp}), and
\bea
&&\phi(x) = \phi_0 - i p_0 \beta x + 
i \sum_{n\ne0} \frac{1}{n} \alpha_n e^{in \beta x} , \nn 
&&\bar{\phi}(x)
= \bar{\phi}_0 + i \bar{p}_0 \beta x +
i \sum_{n\ne0} \frac{1}{n} \bar{\alpha}_n e^{-in \beta x} ,
\eea
where 
\bea
\alpha_{-n} \equiv \sqrt{n} a_n^{\dagger}, &&
\alpha_{n} \equiv \sqrt{n} a_n, \nn
\bar{\alpha}_{-n} \equiv \sqrt{n} \bar{a}_n^{\dagger}, && 
\bar{\alpha}_{n} \equiv \sqrt{n} \bar{a}_n , \quad
(n>0). 
\eea
In the above, we have trivially 
added the Hilbert space for decoupled zero modes,
with appropriate initial and final states for zero modes.
Eq.~(\ref{fermi}) is nothing but the {\it fermionization} formula
familiar in 2D CFT on a circle.
Note that the Fokker-Planck Hamiltonian is different
from the Hamiltonian of the free boson.
Thus, in our model the fermionization of 
the scalar field $\phi(x)$ constructed
from the string creation/annihilation operators, 
can be interpreted as change of variables to D-brane
creation/annihilation operators.
{They are dual variables, in the sense of 2D bosonization}.
The eigenvalue $x$ is interpreted as 
the position of the D-brane on the T-dual circle,
and winding number $n$ becomes momentum in the T-dual picture.
The integration over the position $x$
indicates that this is a quantum description.
In terms of the original AdS$_5$-CFT$_4$ language, 
the D-brane is identified with a D3-brane, or an Euclidean
D2-brane in the T-dual picture.
These D-brane variables are expected to be useful for analyzing
non-perturbative effects of the string theory
on asymptotically
AdS space at finite temperature using the matrix model 
for Polyakov loops. 
See Ref.~\cite{Alvarez-Gaume:2005fv,AlvarezGaume:2006jg} for possible place
for the application. 
See also
Refs.~\cite{Basu:2006mq,Dutta:2007ws} 
for other approaches to the fermionized descriptions 
of the AdS-CFT
correspondence at finite temperature.

In the case of the Hermitian matrix model
\cite{Hanada:2004im,Ishibashi:2007gh},
the counterpart
of the winding number was given by the length of the string.
However, in that case, there appears a sort of
negative string length,
which should be associated with an annihilation of 
string with positive length.
This makes a slight difference compared with our case.
Here, the situation is more symmetric:
The negative winding number just means
that the string winds in the opposite direction
in the Euclidean time.
This difference between the Hermitian matrix model 
and our model
arises from the imaginary number 
$i$ in the exponent of the loop
operators. 
This originates from the fact that
the matrix in the Hermitian models is a ``tachyon", 
whereas
in our case it is a gauge field.
The ways these fields
couple to the boundary of 
the string worldsheet are different.

\section{For CosPA, briefly}

\begin{figure}
\begin{center}
 \leavevmode
 \epsfxsize=90mm
 \epsfbox{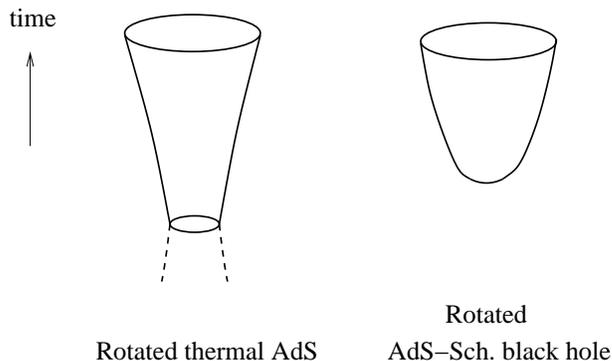}\\
\end{center}
\caption{Interpretation as big bang models.
\label{4cospa}}
\end{figure}

Since this symposium is 
for cosmology and particle astrophysics,
I try to make contact here.
If one interprets the stochastic time
as our real time, i.e. if 
one can Wick rotate the stochastic
time to Lorentzian signature,
the Fig.~\ref{topoc} can be interpreted
as toy model for big bang universe
(Fig.~\ref{4cospa}).
The rotated thermal AdS corresponds to
big bang starting from big crunch,
while 
the rotated AdS-Schwarzschild black hole 
geometry looks like the beginning of the universe
out of nothing \cite{Hawking:1983hj}.
Whether which of these two scenarios
is realized is determined by the 
radius of the Euclidean time circle
(which should now be interpreted as ``space").

Actually, 
it is not so straightforward to make
the Wick rotation of the radial direction
to time direction keeping
the AdS-CFT correspondence \cite{Spradlin:2001nb}.
One may still hope that
our Hamiltonian formulation
may provide better way for 
rotating the time to the Lorentzian signature.

\section{Summary \& future directions}

In this talk,
I have presented a formalism
that translates gauge theory language
into closed string field theory language,
in a simplified setting of the AdS-CFT correspondence.
The deconfined phase in the gauge theory
side is translated into 
the winding string condensation
in the closed string field theory.
The winding string condensation captures
the topological property of the AdS-Schwarzschild black hole
geometry which is dual to the deconfined phase,
in that the winding number is not conserved.
Since at this moment the understanding of the 
field condensation
in closed string field theory
is still limited,
I believe my result in the context of the
AdS-CFT correspondence
provides a useful starting point
for studying the issues
related to the background independence 
in closed string field theory.

In principle, the method I used,
namely the combination of the stochastic quantization
and the change of variables from matrix to loops,
is applicable to the full AdS$_5$-CFT$_4$ correspondence.
It should be the next direction to pursue.

In our setting, the
starting point was the gauge theory,
and we have constructed closed string field theory from
the gauge theory.
As a duality relation, it would be nice to have
an independent definition of the closed string theory.
But also note that there is a possibility that the gauge theory is
more fundamental than the closed string theory,
and if it is the case, 
our approach will 
turn out out to be more fundamental.

I have also shown that the D-brane degrees of freedom
is obtained by fermionization of the string field.
This property is quite special to 
one matrix models,
and points
difficulties in constructing
field theory of D-branes in higher dimensions:
In higher dimensions, when one creates D-brane
one also needs to create open strings
that connect the created D-braes to other D-branes \cite{Yoneya:2007ed}.
We can still hope that some features
in our model will remain useful
even in higher dimensional models.
Indeed, interesting developments have been made
in 26D bosonic string field theory \cite{Baba:2006rs,Baba:2007je,Baba:2007tc}
utilizing the insights obtained
from the experience in non-critical
strings. 
It will be interesting to find 
the dual matrix model of 26D bosonic string field theory, 
if it exists, and apply our method.

\section*{Acknowledgments}

I would like to 
thank the organizers
of the CosPA 2007
for the invitation to present this work.
I would also like to thank 
Masafumi Fukuma,
Takayuki Hirayama,
Pei-Ming Ho,
Nobuyuki Ishibashi,
Toshihiro Matsuo,
Koichi Murakami,
Shunsuke Teraguchi,
Seiji Terashima,
Dan Tomino,
Wen-Yu Wen,
Tamiaki Yoneya
and 
Syoji Zeze
for helpful discussions.
I am also benefited from the visit to KEK and
the String Field Theory 07 symposium at RIKEN.
Part of this work has been carried out 
at my previous 
institutions, Harish-Chandra Research Institute
and Center for Theoretical Sciences at National Taiwan University.
This work was supported in part by National Science Council of Taiwan
under grant No. NSC 96-2119-M-002-001
and NSC 97-2119-M-002-001.

\bibliography{cospa07}
\bibliographystyle{kazu}

\end{document}